\documentclass[a4paper,11pt]{article}
\usepackage{pos}
\usepackage{slashed}

\newcommand{\ba}{\begin{eqnarray}}
\newcommand{\ea}{\end{eqnarray}}

\newcommand{\pkuphy}{School of Physics and State Key Laboratory of Nuclear Physics and Technology, Peking University, Beijing 100871,
	China}

\title{Approach to $K\to \ell\nu_\ell \ell'^+ \ell'^-$ decay width on lattice}

\author*[a]{Xin-Yu Tuo}

\affiliation[a]{\pkuphy}

\emailAdd{ttxxyy@pku.edu.cn}

\abstract{Lattice QCD plays an increasing role in reducing theoretical uncertainties in kaon decays. Here we focus on kaon decay channels $K\to \ell\nu_\ell \ell'^+ \ell'^-$, which are closely related to radiative kaon decays since the lepton pairs $\ell'^+ \ell'^-$ come from virtual photon emission: $K\to  \ell\nu_\ell \gamma^* \to \ell\nu_\ell \ell'^+ \ell'^-$. Compared with real photon emission, these channels involve more complicated form factors due to the off-shell photon with possible large momentum transfer, which causes a challenge to lattice-QCD studies. In this work, we introduce a lattice calculation procedure for their decay width, which can avoid parameterization of form factors. The systematic errors in our method are found to be controllable. Infinite volume reconstruction method is adopted to remove the temporal truncation effects and reduce finite volume effects from kaon intermediate state. This approach sets up a bridge between lattice QCD calculation and experimental measurement of decay width.}

\FullConference{%
 The 38th International Symposium on Lattice Field Theory, LATTICE2021
  26th-30th July, 2021
  Zoom/Gather@Massachusetts Institute of Technology
}

\begin{document}
\maketitle
\section{\label{intro}Introduction}
Kaon decays provide an excellent place for high precision test of Standard Model~\cite{Cirigliano:2011ny}. For this purpose, the theoretical uncertainties in long-distance contribution associated with non-perturbative QCD need to be reduced. In recent years, lattice QCD plays an increasing role on this issue~\cite{Aoki:2021kgd}. An important class is radiative corrections to the leptonic and semileptonic decays, which has also been investigated on lattice recently~\cite{Carrasco:2015xwa,Lubicz:2016xro,Giusti:2017dwk,Feng:2020zdc,Feng:2020mmb,Seng:2020wjq,Desiderio:2020oej,Frezzotti:2020bfa,Seng:2020jtz,Ma:2021azh,Kane:2019jtj}. In this work we consider the decay channels $K\to \ell\nu_\ell \ell'^+ \ell'^-$, where the pair of leptons $\ell'^+ \ell'^-$ is produced by a virtual photon: $K\to \ell\nu_\ell \gamma^*\to \ell\nu_\ell \ell'^+ \ell'^-$. In experiments, three channels have been measured: $K\to e\nu_e e^+ e^-$, $K\to \mu\nu_\mu e^+ e^-$ and $K\to e\nu_e \mu^+ \mu^-$~\cite{Poblaguev:2002ug,Ma:2005iv}. 

Compared with real photon emission, these channels involve more complicated form factors due to the off-shell photon~\cite{Bijnens:1992en}. Since the phase space of $K\to \ell\nu_\ell \ell'^+ \ell'^-$ allows the virtual photon carries relatively large momentum, understanding the momentum dependence of decay amplitude is essential in calculation of the total decay width. This complexity brings difficulties to the lattice-QCD calculation. First, parameterization of form factors for off-shell photon with large momentum transfer is non-trivial. Second, the systematic errors such as finite volume effects or temporal truncation effects, can be enhanced from light intermediate states. One example is that the on-shell $\pi\pi$ intermediate states can cause power-law finite volume effects and exponentially growing contamination in temporal direction.

In this work, we propose a calculation procedure for $K\to \ell\nu_\ell \ell'^+ \ell'^-$ decay width and discuss the systematic errors in our method. For numerical implementation, we refer to our recent paper~\cite{Tuo:2021ewr}. The benefit of our method is to provide a smooth connection between the lattice hadronic matrix element and the decay width without introducing specific parameterization of form factors. The systematic errors from long-distance region, such as temporal truncation effects and finite volume effects, are found to be controllable. We use infinite volume reconstruction (IVR) method~\cite{Feng:2018qpx} to correct errors associated with kaon intermediate state. For $\pi\pi$ intermediate states, we give a qualitative discussion and leave the precise treatment for future works. 

\section{\label{sec2}Calculation Procedure}
In this section, we introduce the calculation procedure for $K\to \ell\nu_\ell \ell'^+ \ell'^-$ decay width. The Euclidean space-time hadronic function $H_E^{\mu\nu}(x,Q)$ in infinite volume is defined as:
\begin{equation}
H_E^{\mu\nu}(x,Q) =\left\langle 0\left|T\left\{J_{e m, E}^{\mu}(x)
J_{W,E}^{\nu}(0)\right\}\right| K(Q)\right\rangle. 
\end{equation}
where $Q=(im_K,\vec{0})$ is the Euclidean 4-momentum of initial kaon state. The electromagnetic and weak currents in Euclidean space are defined as
$J_{em,E}^{\mu}=\frac{2}{3} \bar{u} \gamma^{\mu} u-\frac{1}{3} \bar{d}
\gamma^{\mu} d-\frac{1}{3} \bar{s} \gamma^{\mu} s$ and $J_{W,E}^{\nu}=\bar{s} \gamma^{\nu}\left(1-\gamma_{5}\right) u$. It can be calculated from lattice three-point correlation function: 
\begin{equation}
\label{H}
H_{E,A / V}^{(L),\mu \nu}(x,Q)=
\begin{cases}
N_{K}^{-1} Z_{V} Z_{A / V} e^{m_{K} \Delta T} \left\langle J_{e m}^{\mu}(\vec{x}, t) J_{A/V}^{\nu}(\vec{0}, 0)
\phi_{K}^{\dagger}(-\Delta T)\right\rangle, & t\geq 0, \\
N_{K}^{-1} Z_{V} Z_{A / V} e^{m_{K}(\Delta T-t)} \left\langle J_{A/V}^{\mu}(\vec{0}, 0) J_{e m}^{\nu}(\vec{x}, t)
\phi_{K}^{\dagger}(t-\Delta T)\right\rangle, & t<0.
\end{cases}
\end{equation}
where we add superscript $(L)$ to emphasize that this quantity is defined in finite volume. The axial-vector-current part and vector-current part are separated since they have different renormalization factor $Z_A$ and $Z_V$. The normalization factor
$N_{K}=\langle K|\phi_{K}^{\dagger}(0)| 0\rangle/(2 m_{K})$ and the kaon mass $m_K$ are calculated from the lattice two-point functions. $\Delta T$ is large enough for kaon states dominance.

\begin{figure}[htbp]
	\centering
	\includegraphics[width=0.7\textwidth]{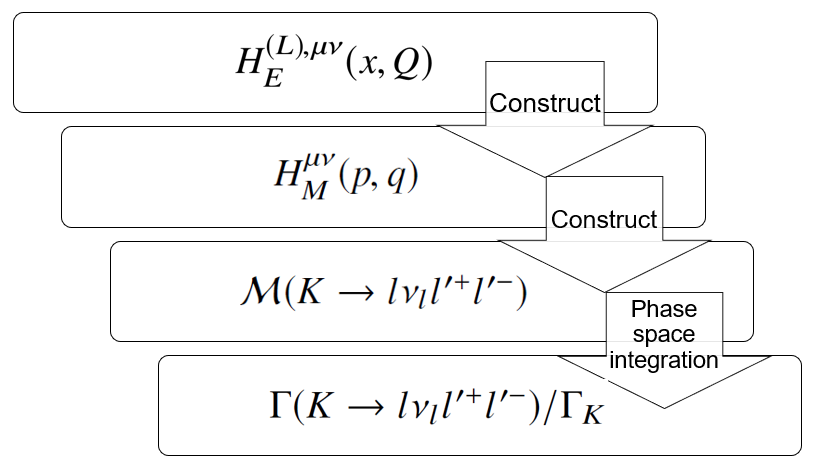}
	\caption{Lattice calculation procedure for $K\to \ell\nu_\ell \ell'^+ \ell'^-$ decay width. \label{fig:pro}}
\end{figure}

Unlike traditional calculation at fixed lattice momenta, we use hadronic function in coordinate space $H_E^{(L),\mu\nu}(x,Q)$ as input. As shown in Fig.~\ref{fig:pro}, we introduce a calculation procedure to provide a direct connection between the lattice hadronic function and the decay width.

In the first step, we relate $H_E^{(L),\mu\nu}(x,Q)$ to Minkowski hadronic function $H_M^{\mu\nu}(p,q)$, which is defined as:
\begin{equation}\label{HM}
H_M^{\mu\nu}(p,q)=\int d^4x\,e^{ipx}\left\langle
0\left|T\left\{J_{em,M}^{\mu}(x)
J_{W,M}^{\nu}(0)\right\}\right| K(q)\right\rangle.
\end{equation}
with Minkowski momenta $p=(E,\vec{p})$, $q=(m_K,0)$. 

We first define hadronic function $H_{E}^{\mu\nu}(P,Q)$ in infinite volume Euclidean space-time: 
\begin{equation}\label{Hcalc}
H_{E}^{\mu\nu}(P,Q)=-i\int_{-T/2}^{T/2} dt\int d^3\vec{x}\, e^{Et-i\vec{p}\cdot\vec{x}}H_E^{\mu\nu}(x,Q).
\end{equation}
where the kinematic variables are $P=(iE,\vec{p})$ and $Q=(im_K,\vec{0})$. On lattice, $H_E^{\mu\nu}(x,Q)$ is replaced by the finite volume version $H_E^{(L),\mu\nu}(x,Q)$:
\begin{equation}\label{HcalcV}
H_{E}^{(L),\mu\nu}(P,Q)=-i\int_{-T/2}^{T/2} dt\int_V d^3\vec{x}\, e^{Et-i\vec{p}\cdot\vec{x}}H_E^{(L),\mu\nu}(x,Q).
\end{equation}

The differences between $H_{E}^{(L),\mu\nu}(P,Q)$ and $H_M^{\mu\nu}(p,q)$ come from two aspects:
\begin{enumerate}
	\item Temporal truncation effects. They account for differences between Eq.~(\ref{HM}) and Eq.~(\ref{Hcalc}), which are caused by non-equivalence between Euclidean space-time and Minkowski space-time.
	\item Finite volume effects. They come from replacing infinite volume $H_E^{\mu\nu}(x,Q)$ by finite volume lattice data $H_E^{(L),\mu\nu}(x,Q)$ in Eq.~(\ref{HcalcV}).
\end{enumerate}

Both effects are found to be controllable through IVR method, which is discussed in Sec.~\ref{sec3} in details.

\begin{figure}[htbp]
	\centering
	\includegraphics[width=0.65\textwidth]{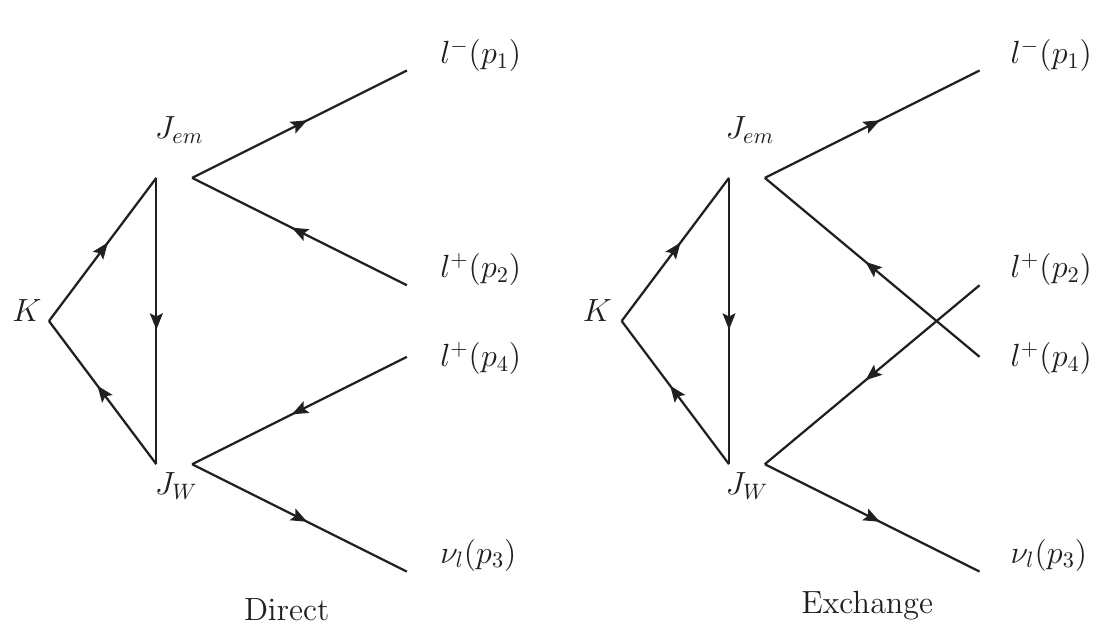}
	\caption{Contribution of off-shell photon radiation from quarks in $K\to
		\ell\nu_\ell \ell^+ \ell^-$.
		The hadronic part is described by $H_M^{\mu\nu}(p,q)$.\label{fig:Fll}}
\end{figure}

In the second step, the decay amplitude $\mathcal{M}(K\to \ell\nu_\ell \ell'^+ \ell'^-)$ is constructed. As shown in Fig.~\ref{fig:IB} and Fig.~\ref{fig:Fll}, it includes two parts: radiation from quarks or from charged lepton ~\cite{Bijnens:1992en}. From these diagrams, the decay amplitude can be written as:
\begin{figure}[htbp]
	\centering
	\includegraphics[width=0.4\textwidth]{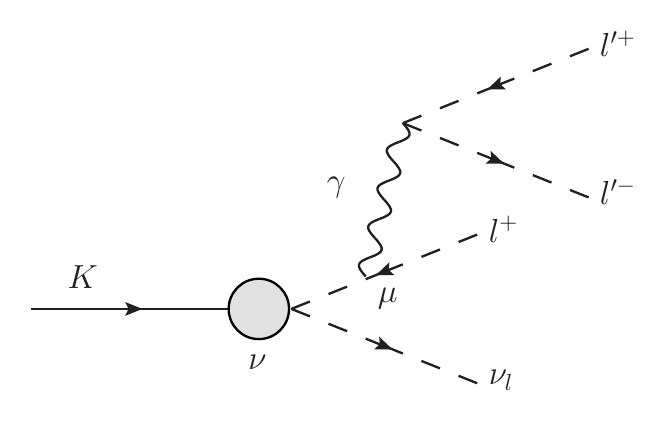}
	\caption{Contribution of off-shell photon radiation from the final-state
		lepton in $K\to \ell\nu_\ell \ell'^+ \ell'^-$. \label{fig:IB}}
\end{figure}

\ba
\label{M}
i \mathcal{M}_{D} &=&-i \frac{G_{F} e^{2} V_{u s}^{*}}{\sqrt{2}
	s_{12}}\left[f_{K} L^{\mu}\left(p_{1}, p_{2}, p_{3}, p_{4}\right)-H_M^{\mu
	\nu}\left(p_{12}, q\right) l_{v}\left(p_{3},
p_{4}\right)\right]\left[\bar{u}\left(p_{1}\right) \gamma_{\mu}
v\left(p_{2}\right)\right],
\nonumber
\\
i \mathcal{M}_{E} &=&+i \frac{G_{F} e^{2} V_{u s}^{*}}{\sqrt{2}
	s_{14}}\left[f_{K} L^{\mu}\left(p_{1}, p_{4}, p_{3}, p_{2}\right)-H_M^{\mu
	\nu}\left(p_{14}, q\right) l_{v}\left(p_{3},
p_{2}\right)\right]\left[\bar{u}\left(p_{1}\right) \gamma_{\mu}
v\left(p_{4}\right)\right].
\ea
where $\mathcal{M}_{D}$ and $\mathcal{M}_{E}$ correspond to ``Direct'' and ``Exchange'' diagrams in Fig.~\ref{fig:Fll}. $p_i$ with $i=1,\cdots,4$ are defined as momenta of final-state leptons. $s_{ij}=p^2_{ij}=(p_i+p_j)^2$ is the momentum square of photon.
The leptonic factors $L^\mu$ and $l^\mu$ are defined as
\begin{eqnarray}
\label{eq:lepton1}
&&l^{\mu}\left(p_{3}, p_{4}\right)=\bar{u}\left(p_{3}\right)
\gamma^{\mu}\left(1-\gamma_{5}\right) v\left(p_{4}\right),
\nonumber\\
&&L^{\mu}\left(p_{1}, p_{2}, p_{3},
p_{4}\right)=l^{\mu}\left(p_{3},p_{4}\right)+{L'}^{\mu}(p_1,p_2,p_3,p_4),
\end{eqnarray}
with
\begin{equation}
\label{eq:lepton2}
{L'}^{\mu}\left(p_{1}, p_{2}, p_{3}, p_{4}\right)=m_{\ell} \bar{u}\left(p_{3}\right)\left(1+\gamma_{5}\right)
\frac{2 p_{4}^{\mu}+\slashed {p}_{12}
	\gamma^{\mu}}{m_{\ell}^{2}-\left(p_{4}+p_{12}\right)^{2}}
v\left(p_{4}\right).
\end{equation}

A convenient way to calculate $\mathcal{M}_{D}$ and $\mathcal{M}_{E}$ is using numerical realization of the spinors and calculating decay amplitude by matrix products. 

In the final step, the decay width is given by four-body phase space integration. For $\ell=\ell'$, both ``Direct'' and ``Exchange'' diagrams contribute:
\begin{equation}
\operatorname{Br}[K\to\ell\nu_\ell\ell^+\ell^-]=\frac{1}{2 m_K \Gamma_{K}}
\int d
\Phi_{4}\,\left(\left|\mathcal{M}_D\right|^{2}+\left|\mathcal{M}_E\right|^{2}+2\operatorname{Re}[\mathcal{M}_D\mathcal{M}^*_E]\right).
\end{equation}

For $\ell\neq\ell'$, we only need ``Direct'' diagram:
\begin{equation}
\operatorname{Br}[K\to\ell\nu_\ell\ell'^+\ell'^-]=\frac{1}{2 m_K \Gamma_{K}}
\int d
\Phi_{4}\,\left|\mathcal{M}_D\right|^{2},
\end{equation}
where $\Gamma_{K}$ is total decay width of kaon. In the practical calculation, we use Monte Carlo method to do phase space integral. Details of four-body phase space as well as numerical implementation can be found in our recent paper~\cite{Tuo:2021ewr}.

\section{\label{sec3}Temporal Truncation Effects and Finite Volume Effects}
\begin{figure}[htbp]
	\centering
	\includegraphics[width=0.6\textwidth]{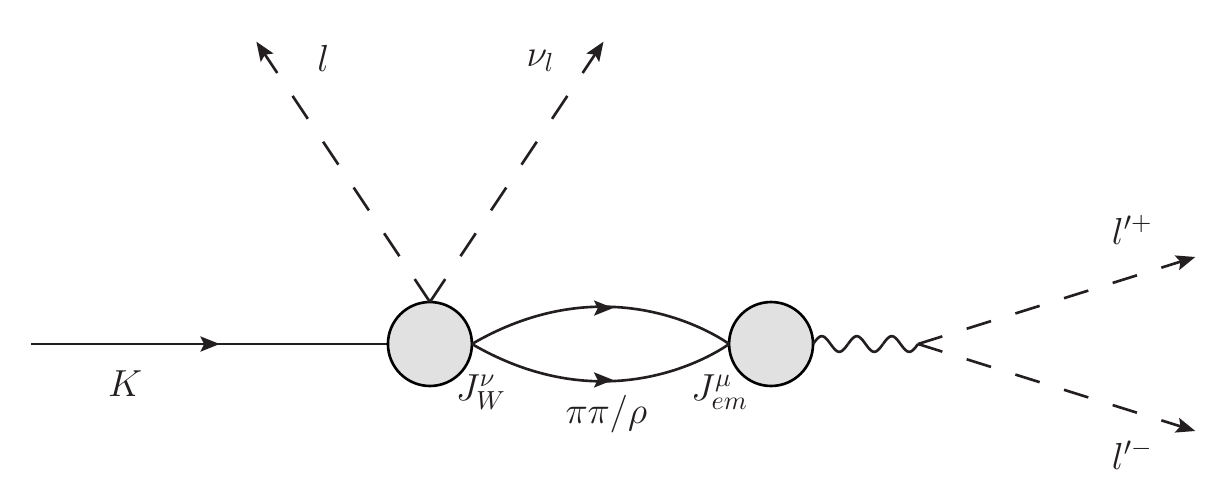}
	\caption{Low-lying-state dominance for $t>0$: $\pi\pi$ or $\rho$ states.\label{fig:pipiint}}
	\includegraphics[width=0.6\textwidth]{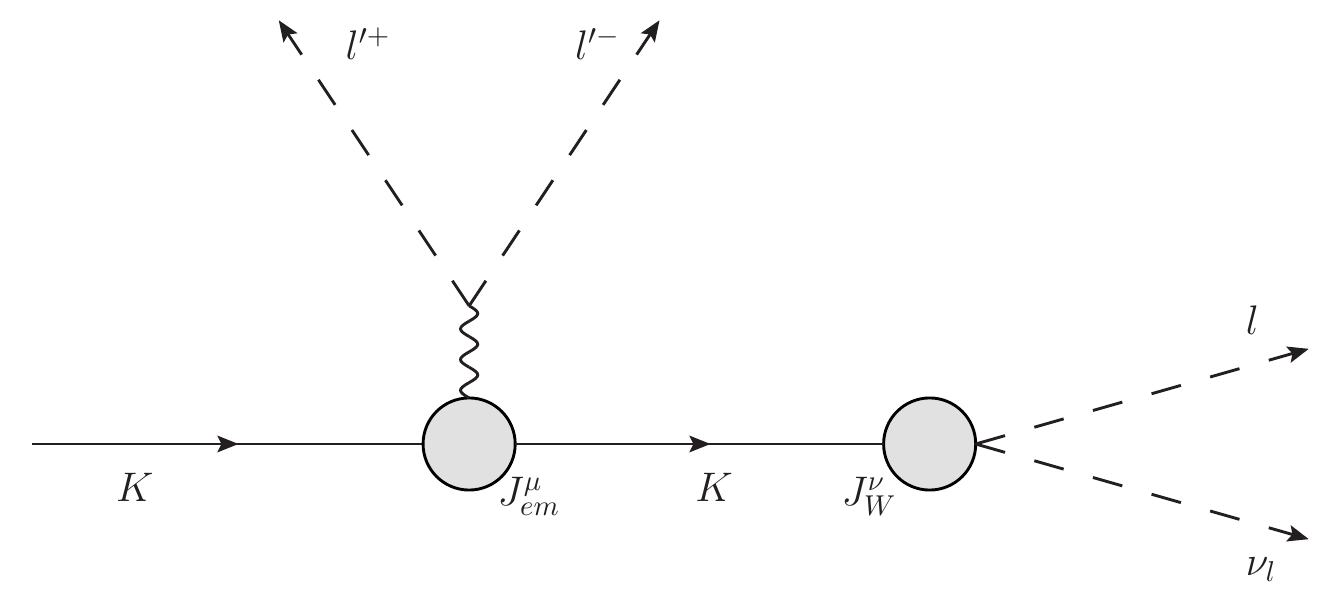}
	\caption{Low-lying-state dominance for $t<0$: kaon states.\label{fig:Kint}}
\end{figure}

In this section, we discuss how to remove temporal truncation effects and reduce finite volume effects. Both effects are dominated by low-lying states. Such states include $K$ intermediate state for $t<0$ time ordering and $\pi\pi$ intermediate state for $t>0$ time ordering, as shown in Fig.~\ref{fig:pipiint} and Fig.~\ref{fig:Kint}. 

\begin{figure}[htbp]
	\centering
	\includegraphics[width=0.5\textwidth]{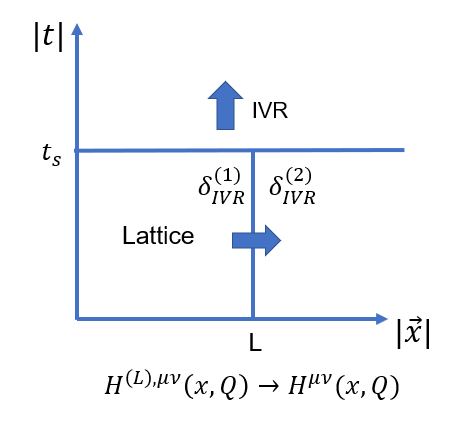}
	\caption{The main idea of IVR method.\label{fig:IVR}}
\end{figure}

IVR method utilizes low-lying-state dominance to reconstruct $H_{E}^{\mu\nu}(P,Q)$ from $H_{E}^{(L),\mu\nu}(P,Q)$. As shown in Fig.~(\ref{fig:IVR}), two steps are included in IVR method: reconstruction in temporal direction (denoted as IVR) and reconstruction in spatial direction (denoted as $\delta_{IVR}$). These two steps correspond to solution of temporal truncation effects and finite volume effects respectively.

We first discuss temporal truncation effects. For $t>0$ and $t<0$ time-ordering, such effects appear as unphysical terms $e^{-(E_{\pi\pi}-E)T/2}$ and $e^{-(E+E_{K}-m_K)T/2}$ respectively. For kaon intermediate state, if the photon is soft and the intermediate kaon is nearly on-shell, $E+E_{K}-m_K\to 0$ will enhance the unphysical term. In relative small lattice $Ta/2=La=2.2fm$, we actually find these effects can not be ignored in the whole phase space~\cite{Tuo:2021ewr}.

This unphysical term is removed by reconstruction in temporal direction. We introduce integration truncation $|t|=t_s$ in temporal direction, where $t_s$ is large enough for kaon state dominance. For $|t|>t_s$, the contribution can be reconstruct from time slice $|t|=t_s$:
\begin{eqnarray}
\label{IVRH}
&&\int^0_{-\infty} dt \int d^3\vec{x}\, e^{Et-i\vec{p}\cdot\vec{x}}H^{\mu\nu}(\vec{x},t)\nonumber\\
=&&\int_{-t_s}^{0} dt \int d^3\vec{x}\,
e^{Et-i\vec{p}\cdot\vec{x}}H^{\mu\nu}(\vec{x},t)+\int_{-\infty}^{-t_s} dt
\int d^3\vec{x}\, e^{Et-i\vec{p}\cdot\vec{x}}H^{\mu\nu}(\vec{x},t)\nonumber\\
=&&\int_{-t_s}^{0} dt \int d^3\vec{x}\,
e^{Et-i\vec{p}\cdot\vec{x}}H^{\mu\nu}(\vec{x},t)+\int d^3\vec{x}\,
e^{-i\vec{p}\cdot\vec{x}}H^{\mu\nu}(\vec{x},-t_s)\frac{e^{-Et_s}}{E+E_K-m_K}.
\end{eqnarray}
From the second to the third line, the hadronic function $H^{\mu\nu}(\vec{x},t)$
with $t<-t_s$ is reconstructed by kaon state dominance. 
Using the hadronic function at some modest value of $t_s$, it
allows us to perform the temporal integral in the whole region of $-\infty <t<-t_s$.
Thus the temporal truncation effects naturally disappear.

For $\pi\pi$ states, the unphysical term is $e^{-(E_{\pi\pi}-E)T/2}$. If $E_{\pi\pi}<m_K$, such states will lead to exponentially growing term. These effects are suppressed by small phase space allowed on-shell two pion $K\to\ell\nu_\ell\pi\pi\to \ell\nu_\ell \ell'^+ \ell'^-$. The correction of this part can be done by reconstruction of low-lying $\pi\pi$ state contribution with $E_{\pi\pi}<m_K$. The treatment of this part will be include in our future work.

Next, we turn to finite volume effects $\delta_{IVR}$. Due to the cluster decomposition property of QCD, $H^{\mu\nu}(x,Q)$ at long-distance is exponentially suppressed as $e^{-\Lambda L}$ with hadronic scale $\Lambda$. Thus, in calculation of decay width, $\delta_{IVR}$ is also exponentially suppressed. They can either be corrected by low-lying state, or be reduced by increasing lattice volume.

We still start with kaon state. Reconstruction in spatial direction is done by:
\begin{eqnarray}\label{DH}
H_E^{\mu\nu}(P,Q)&&=\int d^4 x\, e^{Et-i\vec{p}\cdot\vec{x}}H_E^{\mu\nu}(x,Q)\nonumber\\
&&=\int_V d^4 x\, e^{Et-i\vec{p}\cdot\vec{x}}H_E^{(L),\mu\nu}(x,Q)
\nonumber\\
&&    +\int_V d^4 x\,
e^{Et-i\vec{p}\cdot\vec{x}}\left(H_E^{\mu\nu}(x,Q)-H_E^{(L),\mu\nu}(x,Q)\right)
\nonumber\\
&&+\int_{>V} d^4 x\, e^{Et-i\vec{p}\cdot\vec{x}}H_E^{\mu\nu}(x,Q).
\end{eqnarray}
where the third line comes from the difference between $H_E^{\mu\nu}(x,Q)$ and $H_E^{(L),\mu\nu}(x,Q)$ inside the lattice box, and the fourth line is contribution from $H_E^{\mu\nu}(x,Q)$ outside the lattice box. We denote them as $\delta_{IVR,K}$, which can be calculated using low-lying-state dominance of kaon
\begin{equation}
\label{dIVR2}
\delta_{IVR,K}\approx
\int_V d^3\vec{x}\, e^{Et-i\vec{p}\cdot\vec{x}}(H_K^{(L_\infty),\mu\nu}(x,Q)-H_K^{(L),\mu\nu}(x,Q))
+\int_{>V}d^3\vec{x}\,e^{Et-i\vec{p}\cdot\vec{x}}H_K^{(L_\infty),\mu\nu}(x,Q),
\end{equation}
where we define $L_\infty$ as a much larger lattice volume. The kaon contribution $H_K^{(L),\mu\nu}(x,Q)$ can be calculated analytically with kaon electromagnetic form factor $F^{(K)}(q^2)=1+(r^2_K/6)q^2$:
\begin{equation}
\label{HK}
H_K^{(L),\mu\nu}(x,Q)=\frac{1}{L^3}\sum_{\vec{p}}\frac{1}{ 2E}f_K
P^\nu(P+Q)^\mu F^{(K)}(q^2)e^{-i\vec{p}\cdot \vec{x}}e^{(E-m_K)t}.
\end{equation}

\begin{figure}[htbp]
	\centering
	\includegraphics[width=0.75\textwidth]{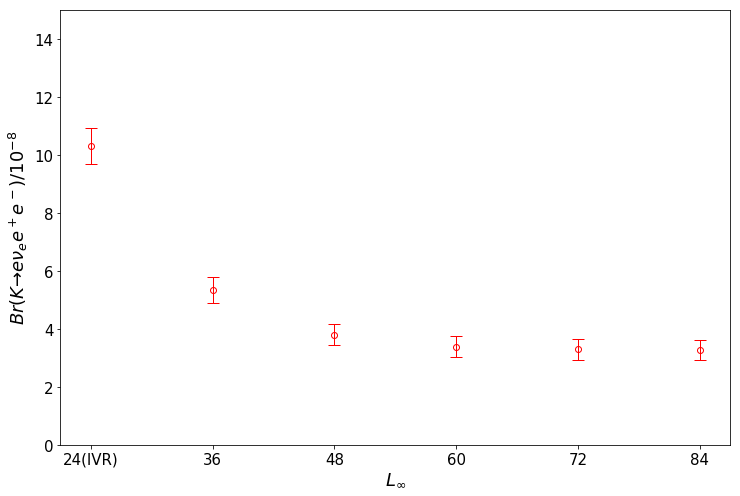}
	\caption{Examination of volume dependence of finite volume effects $\delta_{IVR,K}$ using $K\to e\nu_e e^+e^-$ decay as an example.\label{fig:Linf}}
\end{figure}

As shown in Fig.~\ref{fig:Linf}, $\delta_{IVR,K}$ is exponentially suppressed in calculation of decay width. 

Next, we analyze the $\pi\pi$ intermediate states. There are two sources of finite volume effects. The first one is the power-law finite volume effects from on-shell $\pi\pi$ state. These effects are also suppressed by small phase space $K\to\ell\nu_\ell\pi\pi\to \ell\nu_\ell \ell'^+ \ell'^-$. The second one is the exponentially suppressed finite volume effects $\delta_{IVR,\pi\pi}$, which is similar to kaon case. The correction formulas for both parts will be derived in our future work.

\section{\label{sec4}Conclusion}
This paper presents a methodology for computing the $K\to \ell\nu_\ell \ell'^+ \ell'^-$ decay width using lattice QCD. The calculation procedure is shown in Fig.~\ref{fig:pro}, which uses coordinate space hadronic function as input and does not depend on parameterization of form factors. This approach sets up a bridge between lattice data and experimental observables. For numerical implementation, we refer to our recent paper~\cite{Tuo:2021ewr}.

The systematic errors, including temporal truncation effects and finite volume effects, are found to be controllable in this approach. They are all long-distance effects and are dominated by low-lying states. We adopt IVR method to solve the systematic errors associated with kaon intermediate state. The situation for $\pi\pi$ intermediate state are qualitatively analyzed in this work. The treatment of this part, as well as the other systematic effects, are left for future investigations.

\bibliographystyle{JHEP}
\bibliography{ref}

\end{document}